\documentstyle[12pt,psfig]{article}

\makeatletter
\def\ps@myheadings{\let\@mkboth\@gobbletwo
 \def\@oddhead{\hfil\rightmark}%
 \def\@oddfoot{\hfil\rm\thepage\hfil}%
 \def\@evenhead{\@oddhead}%
 \def\@evenfoot{\@oddfoot}\def\sectionmark##1{}\def\subsectionmark##1{}}
\makeatother

\newcommand{\nc}{\newcommand}

\nc{\be}{\begin{equation}}
\nc{\ee}{\end{equation}}
\nc{\beq}{\begin{equation}}
\nc{\eeq}{\end{equation}}
\nc{\beqa}{\begin{eqnarray}}
\nc{\eeqa}{\end{eqnarray}}
\def\nn{\nonumber}
\def\al{&&\!\!\!\!\!\!\!\!\!}
\def\a{&\!\!\!}

\author{\large  F. Hussain$^1$, R. Iengo$^2$, C. N\'u\~nez$^3$ 
and C. A. Scrucca$^{2,4}$\\ \\
{\normalsize \em $^1$ International Centre for Theoretical Physics, Trieste, 
Italy}\\
{\normalsize \em $^2$ International School for Advanced Studies and INFN, 
Trieste, Italy}\\
{\normalsize \em $^3$ Instituto de Astronom\'{\i}a y F\'{\i}sica del Espacio 
(CONICET), Buenos Aires, Argentina}\\
{\normalsize \em $^{4}$ Institut de Physique Th\'eorique, Universit\'e de 
Neuch\^atel, Switzerland}}

\title{
\vskip -30pt
\normalsize
\begin{flushright}
SISSA REF 140/97/EP
\end{flushright}
\vskip 15pt
\Huge Aspects of D-branes dynamics on orbifolds}

\date{}

\begin{document}

\maketitle

\thispagestyle{myheadings}

\begin{abstract}

We discuss D-brane dynamics in orbifold compactifications of type II 
superstring theory. We compute the interaction potential between two D-branes 
moving with constant velocities and give a field theory interpretation of it 
in the large distance limit.

\end{abstract}

\begin{center}

Talk presented by Claudio A. Scrucca

\end{center}

\section{Introduction and summary}

We study various D-brane configurations in orbifold compactifications which 
are D-particles from the 4-dimensional space-time point of view, but can have 
extension in the compact directions. 
More precisely, the cases of the 0-brane of type IIA and a particular
3-brane of type IIB, turn out to be particularly interesting.

The dynamics of these D-branes is determined by a one loop amplitude 
which can be conveniently evaluated in the boundary state formalism 
\cite{Polcai,Call}. 
In particular, one can compute the force between two D-branes moving with 
constant velocity, extending Bachas' result \cite{Bachas} to 
compactifications breaking some supersymmetry \cite{Hins1}.
Analyzing the large distance behavior of the interaction and its velocity 
dependence, it is possible to read the charges with respect to the massless 
fields, and relate the various D-brane configurations to known solutions 
of the 4-dimensional low energy effective supergravity.

Finally, we discuss the emission of massless particle from two D-branes in 
interaction \cite{Hins2}. We compute the average energy which is radiated when
two D-branes pass each other.

\section{Interactions on orbifolds}

Consider two D-branes moving with velocities $V_1 = \tanh v_1$, 
$V_2 = \tanh v_2$ (say along 1) and transverse positions $\vec Y_1$, 
$\vec Y_2$ (along 2,3).
The potential between these two D-branes is given by the cylinder 
vacuum amplitude and can be thought either as the Casimir energy 
stemming from open string vacuum fluctuations or as the interaction 
energy related to the exchange closed strings between the two 
branes. The amplitude in the closed string channel
\beq
\label{amp}
{\cal A}=\int_{0}^{\infty}dl \sum_s <B,V_1,\vec Y_1|e^{-lH}|B,V_2,\vec Y_2>_s
\eeq
is just a tree level propagation between the two boundary states, which are 
defined to implement the boundary conditions defining the branes. 
There are two sectors, RR and NSNS, corresponding to periodicity and 
antiperiodicity of the fermionic fields around the cylinder, and after
the GSO projection there are four spin structures, R$\pm$ and NS$\pm$,
corresponding to all the possible periodicities of the fermions on the 
covering torus.

In the static case, one has Neumann b.c. in time and Dirichlet b.c. in
space. The velocity twists the 0-1 directions and gives them rotated b.c. 
The moving boundary state is most simply obtained by boosting the static one 
with a negative rapidity $v=v_1-v_2$ \cite{Billo}.
$$
|B,V,\vec Y> = e^{-ivJ^{01}}|B,\vec Y> \;.
$$
In the large distance limit $b \rightarrow \infty$ only world-sheets with
$l \rightarrow \infty$ will contribute, and momentum or winding in the 
compact directions can be safely neglected since they correspond to 
massive subleading components.

The moving boundary states
$$
|B,V_1,\vec Y_1>=\int\frac{d^{3}\vec k}{(2\pi)^{3}}
e^{i \vec k \cdot \vec Y_1} |B,V_1> \otimes|k_B> \;,\;\;
|B,V_2,\vec Y_2>=\int\frac{d^{3}\vec q}{(2\pi)^{3}}
e^{i \vec q \cdot \vec Y_2}|B,V_2> \otimes|q_B> \;,
$$
can carry only space-time momentum in the boosted combinations 
$k_B^\mu = (\sinh v_1 k^1, \cosh v_1 k^1, \vec k_T)$ and $q_B^\mu = 
(\sinh v_2 q^1, \cosh v_2 q^1, \vec q_T)$. 
Notice that because of their non zero velocity, the branes can also transfer
energy, and not only momentum as in the static case.

Integrating over the bosonic zero modes and taking into account momentum 
conservation ($k_B^\mu = q_B^\mu$), the amplitude factorizes into a bosonic 
and a fermionic partition functions:
\beq
{\cal A}=\frac 1{\sinh v} \int_{0}^{\infty}dl 
\int \frac {d^2 \vec k_T}{(2\pi)^2} e^{i \vec k \cdot \vec b} 
e^{-\frac {q_B^2}2} \sum_s Z_B Z^s_F 
=\frac 1{\sinh v} \int_{0}^{\infty} \frac {dl}{2\pi l} e^{- \frac {b^2}{2l}}
\sum_s Z_B Z^s_F \nn 
\eeq
with $Z_{B,F}=<B,V_1|e^{-lH}|B,V_2>^s_{B,F}$. From now on, 
$X^\mu \equiv X^\mu_{osc}$.

It will prove very convenient to group the fields into pairs,
\beqa
X^\pm = X^0 \pm X^1 \a\rightarrow\a \alpha_{n},\beta_{n}= 
a^{0}_{n} \pm a^{1}_{n} \;, \nonumber \\
X^{i},X^{i*} = X^i \pm i X^{i+1} \a\rightarrow\a 
\beta^i_{n},\beta^{i*}_{n}= a^{i}_{n}\pm i a^{i+1}_{n} \;,\;\; i=2,4,6,8  
\;, \nonumber \\
\chi^{A,B} = \psi^0 \pm \psi^1 \a\rightarrow\a \chi^{A,B}_{n}= 
\psi^{0}_{n}\pm\psi^{1}_{n} \;,\nonumber \\
\chi^{i},\chi^{i*}= \psi^i \pm i \psi^{i+1} \a\rightarrow\a \chi^{i}_{n},
\chi^{i*}_{n}= \psi^{i}_{n}\pm i \psi^{i+1}_{n}
\;,\;\; i=2,4,6,8 \;, \nonumber
\eeqa
with the commutation relations $[\alpha_m,\beta_{-n}] = -2 \delta_{mn}$, 
$[\beta_m^i,\beta_{-n}^{i*}] = 2 \delta_{mn}$, 
$\{\chi^A_m,\chi^B_{-n}\} = -2 \delta_{mn}$, 
$\{\chi_m^i,\chi_n^{i*}\} = 2 \delta_{mn}$.
For the RR zero modes, which satisfy a Clifford algebra and are thus 
proportional to $\Gamma$-matrices, $\psi^\mu_o = i \Gamma^\mu /\sqrt{2}$,
$\tilde \psi^\mu_o = i \tilde \Gamma^\mu /\sqrt{2}$,
on can construct similarly the creation-annihilation operators
$$
a,a^* = \frac 12 (\Gamma^0 \pm \Gamma^1) \;,\;\; 
b^i,b^{i*} = \frac 12 (-i\Gamma^i \pm \Gamma^{i+1}) \;,
$$
and similarly for tilded operators, satisfying the usual algebra
$\{a,a^*\} = \{b^i,b^{i*}\} = 1$.

In this way, any rotation or boost will reduce to a simple phase 
transformation on the modes. In fact, for an orbifold rotation ($g_a = 
e^{2\pi i z_a}$) one has
\beqa
\label{orb}
\al \beta _n^a \rightarrow g_a \beta_n^a \;,\;\;
\chi_n^a \rightarrow g_a \chi_n^a \;,\;\;
b^a \rightarrow g_a b^a \;, \nn \\
\al \beta_n^{a*} \rightarrow g^*_a \beta_n^{a*} \;,\;\;
\chi_n^{a*} \rightarrow g^*_a \chi_n^{a*} \;,\;\;
b^{a*} \rightarrow g^*_a b^{a*} \;.
\eeqa
whereas for a boost of rapidity $v$,
\beqa
\label{boost}
\al \alpha_n \rightarrow e^{-v} \alpha_n \;,\;\;
\chi^A_n \rightarrow e^{-v} \chi_n^A \;,\;\;
a \rightarrow e^{-v} a \;, \nn \\
\al \beta_n \rightarrow e^{v} \beta_n \;,\;\;
\chi_n^B \rightarrow e^{v} \chi_n^B \;,\;\;
a^* \rightarrow e^{v} a^* \;.
\eeqa

The boundary state which solves the b.c. can be factorized into a bosonic 
and a fermionic parts; the latter can be further splitted into zero mode and 
oscillator parts, and finally
$$ 
|B> = |B>_B \otimes |B_{o}>_F \otimes |B_{osc}>_F \;.
$$

\subsection{Orbifold construction}

An orbifold compactification can be obtained by identifying points in the 
compact part of space-time which are connected by discrete rotations 
$g = e^{2\pi i \sum_a z_a J_{aa+1}}$ on some of the compact pairs 
$X^a$,$\chi^a$, a=4,6,8. In order to preserve at least one supersymmetry, one 
has to impose $\sum_a z_a = 0$.

We will consider three case: toroidal compactification on $T_6$ ($N=8$ SUSY, 
$z_4=z_6 = z_8 = 0$) and orbifold compactification on $T_2 \otimes T_4/Z_2$ 
($N=4$ SUSY, $z_4= -z_6 = \frac 12$, $z_8 = 0$) and $T_6/Z_3$ ($N=2$ SUSY,
$z_4,z_6 = \frac 13, \frac 23$, $z_8 = -z_4 - z_6$).

The spectrum of the theory now contains additional twisted sectors, in which 
periodicity is achieved only up to an element of the quotient group $Z_N$. 
These twisted states exist at fixed points of the orbifold, and can thus occur
only for 0-branes localized at one of the fixed points. We will not discuss
this case here (see \cite{Hins1}).  

Finally, in all sectors, one has to project onto invariant states to get the 
physical spectrum of the theory which is invariant under orbifold rotations. 
In particular, the physical boundary state is given by the projection
$|B_{phys}>=1/N \sum_k |B,g^k>)$ in terms of the twisted boundary states 
$|B,g^k> = g^k|B>$.

\subsection{0-brane}

Consider first the static case, for which the b.c. are Neumann for time and 
Dirichlet for all other directions (i=2,4,6,8 and a=2,4,6). 
The bosonic b.c. translate into the following equations
\beqa
\al (\alpha_n+\tilde{\beta}_{-n})|B>_B=0 \;,\;\;
(\beta_n+\tilde{\alpha}_{-n})|B>_B=0 \;, \nn \\
\al (\beta^i_n - \tilde\beta^{i}_{-n} )|B>_B=0 \;,\;\;
(\beta^{i*}_n - \tilde\beta^{i*}_{-n} )|B>_B=0 \;, \nn
\eeqa
For the fermions, one has integer or half-integer moding in the RR and NSNS
sectors respectively.
\beqa
\al (\chi^A_n+i\eta \tilde{\chi}^B_{-n})|B_{osc},\eta>_F=0 \;,\;\;
(\chi^B_n+i\eta \tilde{\chi}^A_{-n})|B_{osc},\eta>_F=0 \;, \nn \\
\al (\chi^i_n -i\eta \tilde\chi^{i}_{-n} )|B_{osc},\eta>_F=0 \;,\;\;
(\chi^{i*}_n -i\eta \tilde\chi^{i*}_{-n} )|B_{osc},\eta>_F=0 \nn \;. \\
\al (a+i\eta \tilde a^*)|B_{o},\eta>_F=0 \;,\;\;
(a^*+i\eta \tilde a)|B_{o},\eta>_F=0 \;, \nn \\
\al (b^i -i\eta \tilde b^{i})|B_{o},\eta>_F=0 \;,\;\;
(b^{i*} -i\eta \tilde b^{i*})|B_{o},\eta>_F=0 \;. \nn
\eeqa
Here $\eta = \pm 1$ has been introduced to deal later on with the GSO 
projection.

The boundary state solving these b.c. is easily constructed as a Bogolubov
transformation from a spinor vacuum $|0> \otimes |\tilde{0}>$ defined such 
that $a|0> = \tilde a |\tilde{0}> = b^i|0> = \tilde b^{i*} |\tilde{0}> = 0$.
After applying the boost eq. (\ref{boost}), under which the spinor vacuum
picks up an imaginary phase, $|0> \otimes |\tilde{0}> \rightarrow e^{-v} |0> 
\otimes |\tilde{0}>$, the result is 
\beqa
\label{bs0}
\al |B,V>_B=\exp \left\{\frac{1}{2}\sum_{n > 0}
(e^{-2v}\alpha_{-n} \tilde \alpha_{-n} + e^{2v} \beta_{-n} \tilde \beta_{-n} +
\beta^i_{-n}\tilde\beta^{i*}_{-n} + \beta^{i*}_{-n}\tilde\beta^{i}_{-n})
\right\}|0> \;, \nn \\
\al |B_{osc},V,\eta>_F= \exp \left\{\frac{i\eta}{2}\sum_{n > 0}
(e^{-2v} \chi^A_{-n} \tilde \chi^A_{-n} + e^{2v} \chi^B_{-n} \tilde 
\chi^B_{-n} - \chi^i_{-n}\tilde\chi^{i*}_{-n} - 
\chi^{i*}_{-n}\tilde\chi^{i}_{-n})\right\}|0> \;, \\
\al |B_o,V,\eta>_{RR} = e^{-v} \exp \left\{
-i\eta (e^{2v} a^* \tilde a^* - b^{i*} 
\tilde b^i)\right\}|0> \otimes |\tilde{0}> \;. \nn
\eeqa
The complete boosted boundary state is already invariant under orbifold 
rotations eq. (\ref{orb}). This comes from the fact that the $Z_N$ action 
rotates pairs of fields with the same b.c. and is thus irrelevant.

In both sectors, the fermion number operator reverses the sign of the parameter
$\eta$, that is $(-1)^F|B,V,\eta> = - |B,V,-\eta>$, and the GSO-projected 
boundary state is given by the difference $|B,V> = \frac 12 (|B,V,+>-|B,V,->)$.
There will thus be two kinds of contributions in the amplitude for each sector:
the one with equal $\eta$-parameters for both boundary states and the one with
opposite $\eta$-parameters, giving finally four spin structures.

The partition function can then be computed carrying out some simple oscillator
algebra; the ghosts cancel one untwisted pair, say 2-3, and the result is the 
product of the contributions of the 0-1 pair and the 3 compact pairs.

After the GSO projection, only the three even spin structures R+ and 
NS$\pm$ contribute, and the total bosonic (zero-point energy $q^{-\frac 23}$) 
and fermionic (zero-point energy $q^{- \frac 13}$ for NSNS and 
$q^{\frac 23}$ for RR) partition functions are ($q=e^{-2\pi l}$)
\beqa
\al Z_B = 16 \pi^3 i \sinh v q^{\frac 13} f(q^2)^{4} 
\frac 1{\vartheta_1(i\frac v\pi|2il) \vartheta_1^\prime(0|2il)^3} \;, \\
\al Z_F=q^{-\frac 13} f(q^2)^{-4}
\left\{\vartheta_2(i\frac{v}{\pi}|2il)\vartheta_2(0|2il)^3
-\vartheta_3(i\frac{v}{\pi}|2il)\vartheta_3(0|2il)^3
+\vartheta_4(i\frac{v}{\pi}|2il)\vartheta_4(0|2il)^3\right\} \nn \\
\al \quad \;\;\sim V^4 \;,
\eeqa
corresponding to the usual cancellation of the force between two BPS states
\cite{Polch,Bachas}.
Thus, for the 0-brane we get the same result as the uncompactified theory for
every compactification scheme.

\subsection{3-brane}

Let us now consider a particular 3-brane configuration. In the static case, 
we take Neumann b.c. for time, Dirichlet b.c. for space and mixed b.c. for 
each pair of compact directions, say Neumann for the a directions and 
Dirichlet for the a+1 directions.

The new b.c. for the compact directions are
\beqa
\al (\beta^a_n + \tilde\beta^{a*}_{-n} )|B>_B=0 \;,\;\;
(\beta^{a*}_n + \tilde\beta^{a}_{-n} )|B>_B=0 \;, \nn \\
\al (\chi^a_n +i\eta \tilde\chi^{a*}_{-n} )|B_{osc},\eta>_F=0 \;,\;\;
(\chi^{a*}_n +i\eta \tilde\chi^{a}_{-n} )|B_{osc},\eta>_F=0 \;, \nn \\
\al (b^a +i\eta \tilde b^{a*})|B_{o},\eta>_F=0 \;,\;\;
(b^{a*} +i\eta \tilde b^{a})|B_{o},\eta>_F=0 \;. \nn
\eeqa
Defining a new spinor vacuum $|0> \otimes |\tilde{0}>$ such that
$b^a|0> = \tilde b^{a} |\tilde{0}> = 0$ the compact part of the boundary 
state is constructed in the same way as before. In this case, however, the 
boundary state is not invariant under orbifold rotations, under which the 
modes of the fields transform as in eq. (\ref{orb}) and the new spinor 
vacuum as $|0> \otimes |\tilde{0}> \rightarrow g_a |0> \otimes |\tilde{0}>$.
This was expected since a $Z_N$ rotation now mixes two directions with 
different b.c, and thus the corresponding closed string state does not need 
to be invariant under $Z_N$ rotations.
One finds for the compact part of the twisted boundary state
\beqa
\label{bs3}
\al |B,V,g_a>_B= \exp \left\{-\frac{1}{2}\sum_{n > 0}
(g_a^2 \beta^a_{-n}\tilde\beta^{a}_{-n} + 
g_a^{*2}\beta^{a*}_{-n}\tilde\beta^{a*}_{-n})\right\}|0> \;, \nn \\
\al |B_{osc},V,g_a,\eta>_F=\exp \left\{\frac{i\eta}{2}\sum_{n > 0}
(g_a^{2} \chi^a_{-n}\tilde\chi^{a}_{-n} 
+ g_a^{*2} \chi^{a*}_{-n}\tilde\chi^{a*}_{-n})\right\}|0> \;, \\
\al |B_o,V,g_a,\eta>_{RR} = g_a \exp \left\{-i\eta g_a^{*2} b^{a*} 
\tilde b^{a*} \right\}|0> \otimes |\tilde{0}> \;. \nn
\eeqa

After the GSO projection, the total partition functions for a given relative
angle $w_a$ are
\beqa
\al Z_B=16 i \sinh v q^{\frac 13} f(q^2)^4 
\frac 1{\vartheta_1(i \frac v\pi|2il)} 
\prod_a \frac {\sin \pi w_a}{\vartheta_1(w_a|2il)} \;, \\
\al Z_F=q^{-\frac 13}f(q^2)^{-4}
\left\{\vartheta_2(i\frac{v}{\pi}|2il)\prod_a \vartheta_2(w_a|2il) 
\right. \nn \\ \al \qquad \qquad \qquad \qquad \; \left.
-\vartheta_3(i\frac{v}{\pi}|2il)\prod_a \vartheta_3(w_a|2il)
+\vartheta_4(i\frac{v}{\pi}|2il)\prod_a \vartheta_4(w_a|2il)\right\} \nn \\
\al  \quad \;\; \sim \left\{
\begin{array}{l}
V^4 \;\;,\;\;w_a=0 \\
V^2 \;\;,\;\;w_a \neq 0 
\end{array}
\right. \;.
\eeqa
Recall that to obtain the invariant amplitude, one has to average over all 
possible angles $w_a$.

\section{Large distance limit}

In the large distance limit $l \rightarrow \infty$, explicit results with
exact dependence on the rapidity can be obtained and compared to 
a field theory computation. One finds the following behaviors:

\noindent
{\bf 0-brane}
\beq
{\cal A} \sim 4 \cosh v - \cosh 2v - 3 \sim V^4 \;.
\eeq

\noindent
{\bf 3-brane}
\beqa
\al {\cal A}(w_a) \sim 4 \prod_a \cos \pi w_a \cosh v - \cosh 2v - 
\sum_a \cos 2 \pi w_a \;, \nn \\
\al {\cal A} \sim \left\{
\begin{array}{l}
\cosh v - \cosh 2v \sim V^2 \;\;,\;\; T_6/Z_3 \\
4 \cosh v - \cosh 2v - 3 \sim V^4\;\;,\;\; T_2 \otimes T_4/Z_2 \;,\; T_6
\end{array}
\right. \;.
\eeqa

\noindent
The additional twisted sectors can be analyzed similarly, and one finds 
${\cal A} \sim \cosh v - 1 \sim V^2$.

In the low energy effective supergravity field theories, the possible 
contributions to the scattering amplitude in the eikonal approximation come
from vector exchange in the RR sector and dilaton and graviton exchange in 
the NSNS sector. 
The respective contributions have a peculiar dependence on the 
rapidity reflecting the tensorial nature and are: 
\beq
{\cal A}^{NS}_{\phi} \sim -a^2 \;,\;\;
{\cal A}^{R}_{V_\mu} \sim e^2 \cosh v \;,\;\;
{\cal A}^{NS}_{g_{\mu\nu}} \sim -M^2 \cosh 2v \;.
\eeq

Thus, the interpretation of the behaviors found in the various sectors
and for the various brane configurations we have considered, is the following:
\beqa
4 \cosh v - \cosh 2v - 3 \quad \a \Leftrightarrow \a \quad 
\mbox{$N=8$ Grav. multiplet} \;, \nn \\   
\cosh v - \cosh 2v \quad \a \Leftrightarrow \a \quad
\mbox{$N=2$ Grav. multiplet} \;, \nn \\
\cosh v - 1 \quad \a \Leftrightarrow \a \quad 
\mbox{Vec. multiplet} \;. \nn
\eeqa

The patterns of cancellation suggest that all the D-brane configurations
that we have considered correspond to extremal 0-brane solutions of the 
low energy 4-dimensional supergravity, possibly coupling to additional twisted
vector multiplets; the 3-brane configuration on the $Z_3$ orbifold seems to be
an exception since it does not couple to scalars, and should thus correspond 
to a Reissner-Nordstr\"om extremal black hole. 

Finally, notice that $V^2$ terms in the effective action give a non flat
metric to the moduli space. Since in the dual open string channel a constant
velocity $V$ corresponds by $T$-duality to a constant electric field $E$, 
$V^2$ terms correspond to a renormalization of the Maxwell term $E^2$.
It is well known that this can not happen for maximally supersymmetric 
theories; the $V^2$ behavior is thus forbidden for $N=8$ compactifications,
but generically allowed for compactifications breaking some supersymmetry,
$N < 8$. Our results are compatible with this and show that $V^2$ terms do 
indeed appear in some cases.

\section{Particle emission}

For non zero relative velocity between the branes, particle emission is
kinematically allowed even in the eikonal approximation. The corresponding 
amplitude can be computed inserting the appropriate vertex operator in the
matrix element (\ref{amp}); we have done it for various NSNS particle emission
\cite{Hins2}.
At large inter-brane distances, emission occurs only for the scalars of the 
$N=8$ gravitational multiplet (hence there is no scalar emission in the case
of the 3-brane which couples to the $N=2$ gravitational multiplet) and for
the 4-dimensional graviton. One can compute the average energy radiated 
through graviton emission, finding
\beq
<p> \sim g_s^2 l_s^2 \frac {V^{1 + 2n}}{b^3} \;,
\eeq
with $n=2,4$ depending on the amount of preserved supersymmetry. 

\vskip 20pt
\noindent
{\Large \bf Acknowledgments}
\vskip 10pt

Work partially supported by EEC contract ERBFMRX-CT96-0045.


\begin{thebibliography}{99}

\bibitem{Polcai} J. Polchinski and T. Cai, {\em Nucl. Phys.} {\bf B296} 
(1988) 91.

\bibitem{Call} C.G. Callan, C. Lovelace, C.R. Nappi and S.A. Yost, 
{\em Nucl. Phys.} {\bf B293} (1987) 83; Nucl. Phys. {\bf B308} (1988) 221.

\bibitem{Bachas} C. Bachas, {\em Phys. Lett.} {\bf B374} (1996) 49.

\bibitem{Hins1} F. Hussain, R. Iengo, C. N\'u\~nez and C.A. Scrucca,
{\em Phys. Lett.} {\bf B409} (1997) 101.

\bibitem{Hins2} F. Hussain, R. Iengo, C. N\'u\~nez and C.A. Scrucca {\it
``Closed string radiation from moving D-branes''}, hep-th/9710049.

\bibitem{Billo} M. Bill\'o, P. Di Vecchia and D. Cangemi,
Phys. Lett. {\bf B400} (1997) 63.

\bibitem{Polch} J. Polchinski, {\em Phys. Rev. Lett.} {\bf 75} (1995) 4724.

\end{thebibliography}
\end{document}